\documentclass[12pt]{article}

\usepackage{amsmath,amssymb,setspace,graphicx,url,enumitem,soul,float}

\newcommand{\bs}{\boldsymbol}

\usepackage{authblk}
\author[1,2,3]{Li-Chun Zhang}
\affil[1]{\em \small University of Southampton (email: L.Zhang@soton.ac.uk)}
\affil[2]{\em \small Statistics Norway}
\affil[3]{\em \small University of Oslo}
\title{\Large Proxy expenditure weights for Consumer Price Index:\\ Audit sampling inference for big data statistics}
\date{}

\begin{document}

\maketitle

\noindent
\emph{Abstract:} Purchase data from retail chains provide proxy measures of private household expenditure on items that are the most troublesome to collect in the traditional expenditure survey. Due to the sheer amount of proxy data, the bias due to coverage and selection errors completely dominates the variance. We develop tests for bias based on audit sampling, which makes use of available survey data that cannot be linked to the proxy data source at the individual level. However, audit sampling fails to yield a meaningful mean squared error estimate, because the sampling variance is too large compared to the bias of the big data estimate. We propose a novel accuracy measure that is applicable in such situations. This can provide a necessary part of the statistical argument for the uptake of big data source, in replacement of traditional survey sampling. An application to disaggregated food price index is used to demonstrate the proposed approach.


\section{Introduction}

One of the most important uses of the Consumer Expenditure Survey (CES) is to provide the expenditure shares (or weights) for the Consumer Price Index (CPI). However, the CES has a relatively large sampling variance due to its limited sample size, which does not permit disaggregation of CPI for sub-populations that are often of considerable interest. Moreover, it is extremely burdensome, especially due to the diary component, has a very high nonresponse rate, and is known to suffer from various misreporting errors; see e.g. Frickr et al. (2015), Battistin and Padula (2016), and Bee et al. (2012).

Purchase data from retail chains provide \emph{proxy} measures of household expenditure on the items that are most troublesome to collect in the survey, yielding \emph{proxy CPI weights} based on the corresponding expenditure shares. Unlike the CES-based weights, the proxy weights can be considered to have virtually zero sampling variance for practical purposes due to the sheer amount of data that can be made available. However, these weights are biased due to (a) coverage errors caused by the discrepancy between the purchases at the reporting retail chains and the target universe of household consumption, and (b) selection errors from the reported purchases because, for various technical reasons, one is only able to classify and make use of a subset of all the products in the reported purchases. 

In a situation where the proxy weights bias completely dominates the variance, modelling the variability of proxy weights would be fruitless, as long as it cannot capture the inherent bias. Additional observations of expenditure are necessary to investigate whether or to which extent the proxy weights may be biased. The thrust of this paper is to develop an \emph{audit sampling} approach to the following two relevant questions in this context:
\begin{itemize}[leftmargin=4mm]
\item[]{\bf I.} How to \emph{test} the potential bias resulting from the proxy CPI weights?

\item[]{\bf II.} How to \emph{measure} the accuracy of the price index based on proxy CPI weights? 
\end{itemize}

We raise these two questions from the perspective that big transaction data can possibly \emph{replace} the CES altogether, in routine production of the CPI for the relevant consumption sub-universe, but confidence and accuracy measures are required to provide the statistical argument for doing so. Hence, the available CES data will be treated as an audit sample. Such a perspective has certain fundamental differences to the survey sampling outlook and the traditional application of statistical techniques for auditing.

First, there is a longstanding tradition of survey sampling for the estimation of finite population parameters, where auxiliary information can be used to improve the efficiency, by appropriate weighting adjustment or prediction modelling; see e.g. S\"{a}rndal et al. (1992) and Valliant et al. (2000), respectively. One can approach the CES from such a perspective, where the target parameters are the yearly expenditure shares of the private households in a country, and the purchase data that yield the proxy CPI weights are relevant auxiliary information. Nevertheless, we are not aware of any existing practice where the two sources are combined in this way. A major obstacle is privacy concern, against linking individual observations in the CES to the purchase data from the retailers. Another is the extra cost and response burden required to collect the relevant data from the businesses. 

Next, audit sampling techniques for Accounting (e.g. Neter, 2011) can be relevant, if one treats the expenditure measures derived from purchase data as the \emph{book (proxy) amounts}, and use sampling from the population of these proxy measures to obtain a sample of \emph{audited (correct) amounts}, e.g. in order to estimate the error of book amount total and to analyse the individual book amount errors. However, this would require taking a sample from the purchase data directly, which is not feasible due to the same objections above against privacy, cost and burden. In contrast, under our approach, the CES data constitute a \emph{separate} sample, which cannot be linked to the purchase data at the individual level. 

The rest of paper is organised as follows. In Section \ref{data}, we introduce the data for this study and the objective of disaggregated CPI motivated by the available data. In Section \ref{testing} we develop tests for the source effects arising from replacing the CES-based weights by the purchase data proxy weights, in relation to question-I. In Section \ref{uncertainty}, we develop a novel accuracy measure based on audit sampling for big data statistics, whose bias completely dominates the variance, in relation to question-II. We apply the measure to disaggregated food price index, and contrast it to the use of mean squared error. A short summary of some final remarks are given in Section \ref{final}.

\section{Data and disaggregated CPI} \label{data}

We introduce here the data for this study, and the objective of disaggregated CPI motivated by the available data. This provides the context for an age-group-specific food price index, and the application of the audit sampling approach to be developed.

\begin{figure}[ht] 
\resizebox{163mm}{134mm}{\includegraphics{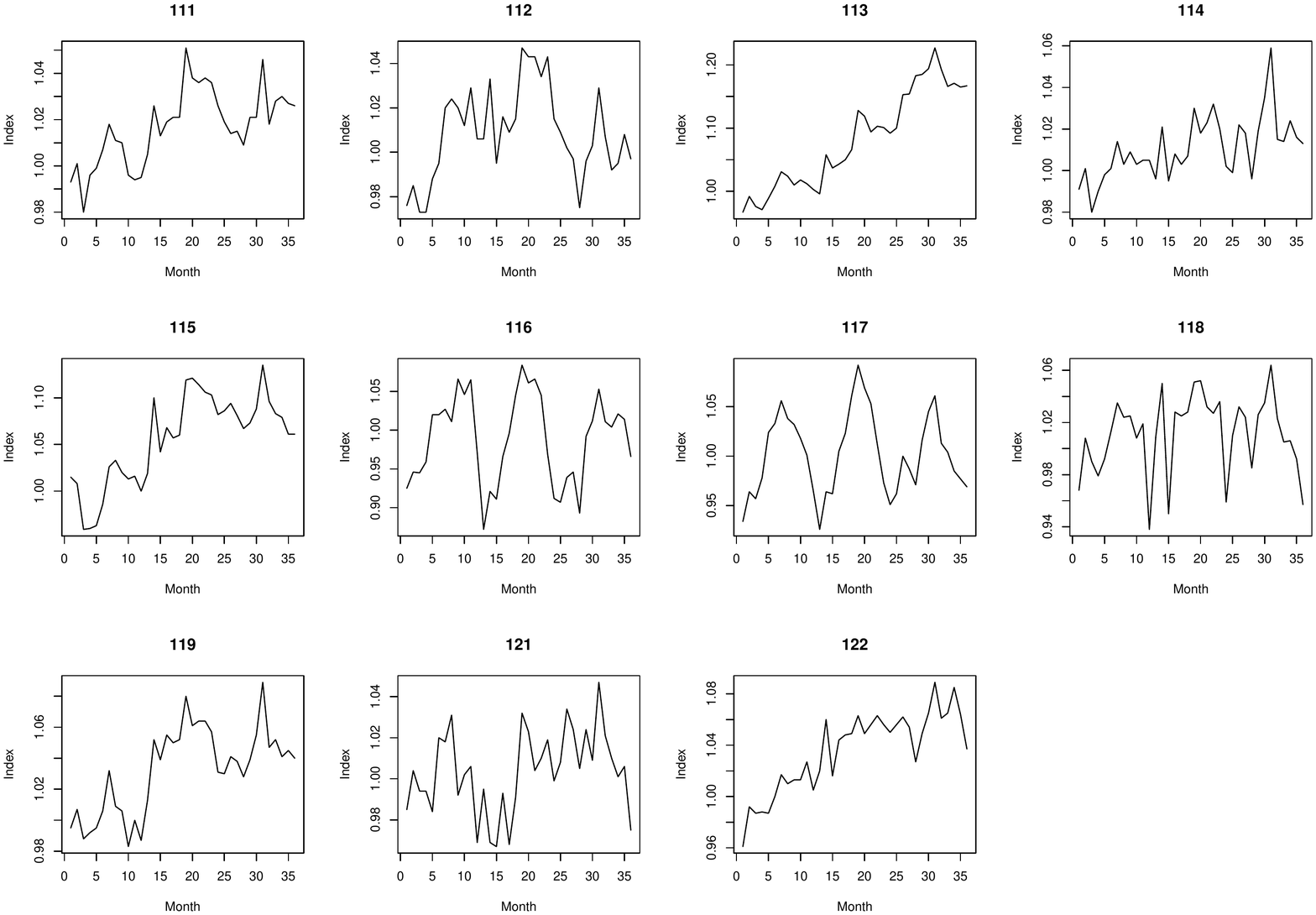}}
\caption{COICOP group price index over 36 months (Source: Statistics Norway)} \label{fig-index}
\end{figure}

The most burdensome CES diary component concerns food and drinks. According to the classification of individual consumption by purpose (COICOP), developed by the United Nations Statistics Division and adopted by the Eurostat, these pertain to the 3-digit COICOP groups 111 - 119, 121 and 122. Figure \ref{fig-index} shows these 11 indices over 36 months of 2015 - 2017, denoted by $T=36$, published by Statistics Norway. One can detect seasonal patterns and/or trends more readily in some of the groups than the others. 

The weights used for these indices are not based on the CES, which was discontinued after 2012. Instead they are proxy weights compiled based on retailer turnovers which are available to the System of National Accounts. A drawback with this \emph{supplier-data approach} is lack of disaggregation over consumer demography, so that one cannot e.g. calculate a CPI for households of pensioners which clearly would be of interest. 

While the CES in principle can facilitate such disaggregation of the CPI, in practice one is often not able to overcome the limitations imposed by the overall CES sample size. Throughout the first decade of this millennium, the sample size of the Norwegian CES is about 7000 households, and the response rate is about 50\%. The diary data are collected over a period of two weeks for each respondent household. For this study, we shall ignore the nonresponse effects and only calculate the sampling variances of the CES-based CPI weights, as if the respondent households were a simple random sample by design.

Table \ref{tab-ces} gives the CES-based weights at the COICOP 2-digit level in 1998-2000 and 2012. The 11 groups in Figure \ref{fig-index} together constitute the first 2-digit category Food and non-alcoholic drinks. It can be seen that their total expenditure share decreased by only 0.2\% over the decade, which on average is less than 0.02\% each year. For disaggregation of the CPI by age, one needs to at least break down each yearly change to the 11 types of goods and for consumers of different age groups. Clearly, the CES respondent sample size is simply too small to support such a high demand of information. 

\begin{table}[ht]
\centering
\caption{Household expenditure, total in NOK. (Source: ssb.no)}
\begin{tabular}{| l |r|r|r|r|} \hline
& \multicolumn{2}{c|}{1998 - 2000} & \multicolumn{2}{c|}{2012} \\ \cline{2-5}
& Total & $\%$ & Total & $\%$ \\ \hline 
\textit{Consumption in all} & {\em 280078} & {\em 100} & {\em 435507} & {\em 100} \\ 	
01 Food, non-alcoholic drinks    & 33499	& 12,0 & 51429 & 11,8\\
02 Alcohol, tobacco & 8114 & 2,9 & 11717 & 2,7\\
03 Clothing, shoes & 16278 & 5,8 & 23618 & 5,4\\
04 Housing, household energy & 71278 & 25,4 & 135982 & 31,2\\
05 Furniture, household articles & 17321 & 6,2 & 24495 & 5,6\\
06 Health & 7717 & 2,8 & 11421 & 2,6\\
07 Transport & 56832 & 20,3 & 81574 & 18,7\\
08 Post, telecommunication & 5610	 & 2,0 & 8253 & 1,9\\
09 Culture, recreation & 33634	& 12,0 & 43347 & 10,0\\
10 Education & 869 & 0,3 & 985 & 0,2 \\
11 Restaurant, hotel, etc. & 11379 & 4,1 & 15557 & 3,6\\
12 Other goods or services & 17547 & 6,3 & 27129 & 6,2 \\ \hline 
\end{tabular} \label{tab-ces} \end{table}

Now, the supplier-based proxy CPI weights can be disaggregated provided it is possible to connect the who (i.e. consumer) and what (i.e. goods) of the purchases. One possibility is via the retailer loyalty members, whose membership IDs are registered together with the purchases. There are obviously issues regarding the coverage of loyalty members and their registered purchases, the confidentiality restrictions and the burden on the businesses. An alternative venue is via the card transactions, as an increasingly overwhelming proportion of all purchases are paid by card (or other digital means), and a handful payment services account for most of the transactions. The separate data of purchases and card transactions can be deterministically compared in terms of non-personal keys such as time, place/outlet and amount, connecting thus the cardholder to the linked purchases. Standard disclosure control methods can be applied to preserve data confidentiality. For disaggregation of the CPI by age, expenditure data by age groups can be extracted from the linked data, without revealing the cardholder identity or the time/place of purchases.

For this study, we have fully anonymised expenditure data based on extractions provided by the largest card payment service and some of the largest retail chains in Norway. The data pertain to a single weekday in September of 2016, consisting of 0.8 million transactions, broken down in four age groups of the cardholder: up to 25, 26 - 40, 41 - 67, 68 and above, denoted by $g = 1, ..., 4$. More data can be acquired if the proxy weights are actually to be used in routine production. Thus, in this study one may consider the proxy weights to have practically zero variance compared to the sampling variance of the CES weights. 

To investigate question-I raised in the Introduction, we use the last CES in 2012 as an independent audit sample, where the expenditure data are broken down by the age of household head in the same four groups. In light of the household demography in Norway, had the two datasets referred to the same time point, one could have expected a large overlap of the eldest age group, i.e. $g=4$, in the two sources. Whereas the overlap will be relatively smaller between the other age-groups, i.e. $g=1, 2, 3$.

\section{Testing proxy source effects} \label{testing}

\subsection{Proxy source effect}

Without losing generality, the CPI for a given universe of commodities, such as food and non-alcoholic drinks (or simply food) in this paper, can be generically given by 
\[
P = \sum_{i=1}^m w_i p_i 
\] 
where $p_i$ is the price index for the $i$-th consumption group, for $i=1, ..., m$, and $w_i$ is the corresponding true expenditure share, where $\sum_{i=1}^m w_i = 1$. Notice that $w_i$ here refers to the expenditure of people in all ages; a subscript $g$ will be added later on when we come to age-group-specific weights and indices.  

Denote by $\hat{w}_i$ the weight estimated from the CES. Denote by $w_i^*$ the big data proxy weight. For the purpose of this paper, we shall assume the CES-weights to be unbiased, denoted by $E(\hat{w}_i) = w_i$ for all $i =1, ..., m$, but are subjected to sampling errors; whereas the proxy weights have zero variances but are biased generally speaking, denoted by $w_i^* \neq w_i$. The price index methodology is undergoing substantive developments regarding the use of scanner data, which is itself a big data source; see in particular the website of Ottawa group (\url{http://www.ottawagroup.org}). For our purpose here, we shall treat the price indices as given, thereby avoiding any associated methodological issues and uncertainty.

Let $\bs{w} = (w_1, ..., w_m)^{\top}$, $\hat{\bs{w}} = (\hat{w}_1, ..., \hat{w}_m)^{\top}$ and $\bs{w}^* = (w_1^*, ..., w_m^*)^{\top}$. Let $\bs{p} = (p_1, ..., p_m)^{\top}$. The observed \emph{source effect} on the CPI due to replacing $\hat{\bs{w}}$ by $\bs{w}^*$ can be given as
\begin{equation}\label{source}
\Delta = P(\hat{\bs{w}}) - P(\bs{w}^*) = \bs{p}^{\top} (\hat{\bs{w}} - \bs{w}^*)  
= \sum_{i=1}^m w_i^* b_i p_i = Cov(b_i, p_i; \bs{w}^*) ~, 
\end{equation}
where $b_i = \hat{w}_i/w_i^* -1$ is the relative difference between the survey and proxy weights, and $Cov(\cdot, \cdot; \bs{w}^*)$ denotes covariance calculated with respect to $\bs{w}^*$ when $\bs{w}^*$ is considered as a probability mass function. The last equality in \eqref{source} follows since, given whichever $\bs{p}$ and $\hat{\bs{w}}$, the mean of $b_i$ with respect to $\bs{w}^*$ is always equal to 0, i.e.
\[
E(b_i; \bs{w}^*) = \sum_{i=1}^m w_i^* b_i \equiv 0 ~.
\] 

The expression \eqref{source} is important conceptually, because it shows that one does \emph{not} need to have $\hat{\bs{w}} = \bs{w}^*$, in order to avoid pronounced source effect $\Delta$, as long as the relative weight difference $\hat{w}_i/w_i^* -1$ and the group-index fluctuation $p_i - P(\bs{w}^*) = p_i - E(p_i; \bs{w}^*)$ are uncorrelated with respect to $\bs{w}^*$ in the sense of 
\[
Cov(b_i, p_i; \bs{w}^*) =0 ~.
\] 
That is, whether or not a group has a higher than average price index is not related to whether or not its CES-based weight is higher than the corresponding proxy weight. 

Regardless if the observed source effect $\Delta$ seems negligible for practical purposes, the related question of inference is whether the true or \emph{expected} source effect is zero. One may rephrase this objective in terms of the following hypothesis test
\begin{equation} \label{H0H1}
H_0:~ E(\Delta) = 0~ \text{ vs. }  H_1:~ E(\Delta) \neq 0 ~,
\end{equation}
where
\[
E(\Delta) = \bs{p}^{\top} \bs{w} - \bs{p}^{\top} \bs{w}^* = P(\bs{w}) - P(\bs{w}^*) 
\qquad\text{and}\qquad V(\Delta) = \bs{p}^{\top} V(\hat{\bs{w}}) \bs{p} ~.
\]
Provided normal distribution of $\bs{p}^{\top} \hat{\bs{w}}$, which is reasonable given the CES sample size, one obtains immediately a standard test statistic
\begin{equation}\label{Ztest}
Z = \Delta/\sqrt{V(\Delta)} ~\sim~ N(0,1) ~.
\end{equation}

\subsection{Application of $Z$-test}

\begin{figure}[ht] 
\hspace{10mm} \resizebox{143mm}{114mm}{\includegraphics{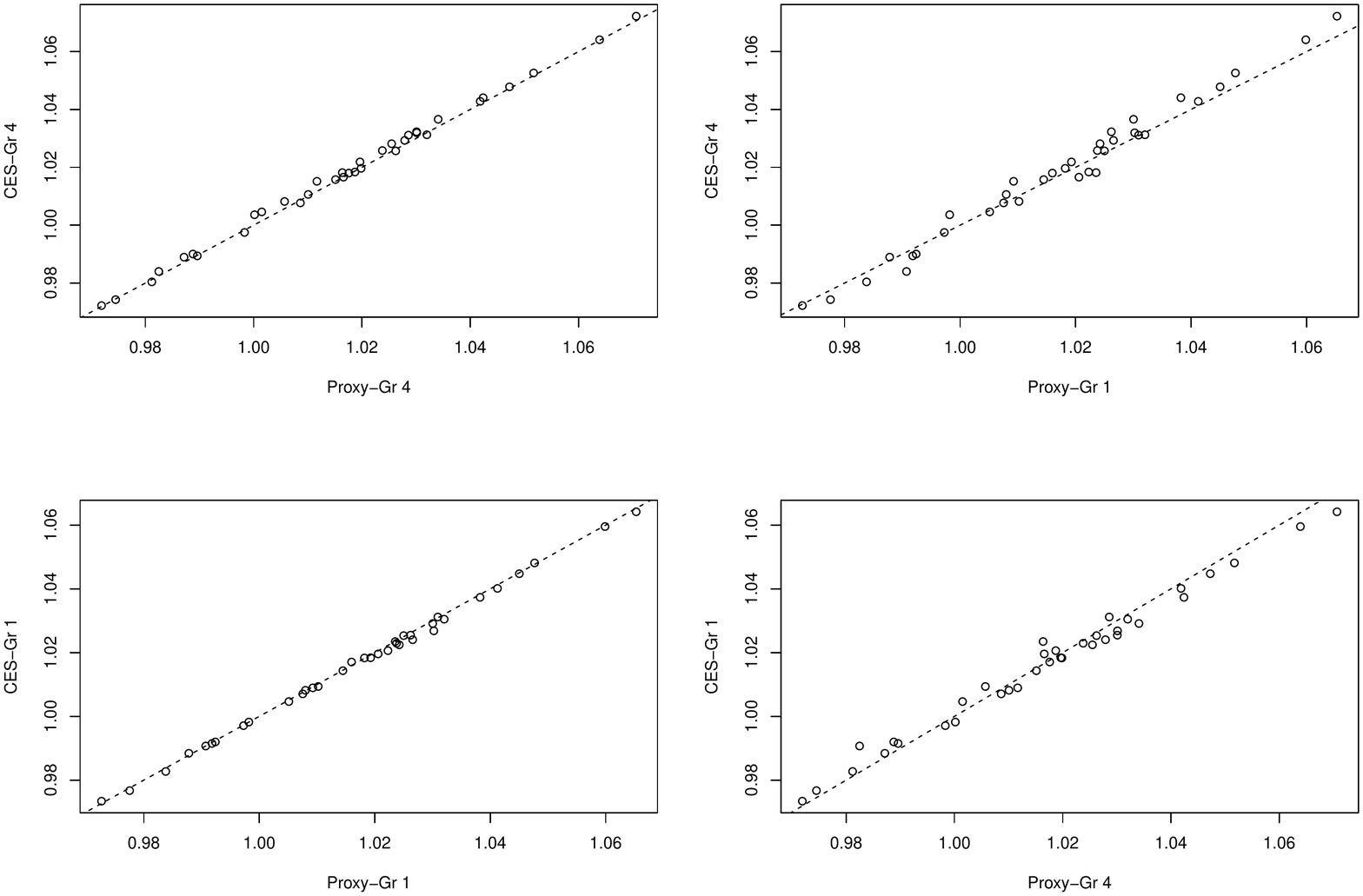}}
\caption{Scatter plots of age-group (Gr) index, CES-based vs. proxy weights.} \label{fig-scat}
\end{figure}

Let $\widehat{P}_{gt} = \bs{p}_t^{\top} \hat{\bs{w}}_g$ be the $g$th age-group index in month $t$, for $g = 1, ..., 4$ and $t=1, ...,T$, where $\bs{p}_t = (p_{1t}, ..., p_{mt})^{\top}$ and $\hat{\bs{w}}_g$ are the CES-based weights for $g$th age-group $g$. Similarly, let $P_{gt}^* = \bs{p}_t^{\top} \bs{w}_g^*$ be the age-group monthly index based on the proxy weights. Figure \ref{fig-scat} shows four scatter plots of $\widehat{P}_{gt}$ vs. $P_{g' t}^*$, where $(g,g') =$  (4,4), (4,1), (1,1) and (1,4). A pair of indices are calculated for two consumer groups of closer population proximity if $g = g'$. As can be seen, $\widehat{P}_{4t}$ and $P_{4t}^*$ for the fourth age-group (top-left in the figure) appear to be scattered around the unity slope (dashed line), as well as $\widehat{P}_{1t}$ and $P_{1t}^*$ for the first age-group (bottom-left). In contrast, the points $(P_{1t}^*,\widehat{P}_{4t})$ appear to scatter around a different slope to the unity (top-right). Similarly for $\widehat{P}_{1t}$ vs. $P_{4t}^*$ (bottom-right).

To apply the $Z$-test \eqref{Ztest} to the $g$th age-group index of a particular month $t$, let
\[
\Delta_t = \widehat{P}_{gt} - P_{gt}^* 
\]
be the observed source effect in month $t$. Or, one can use the observed mean source effect over a number of months, say,
\[
\Delta = \widehat{\bar{P}}_g - \bar{P}_g^* = \bar{\bs{p}}^{\top} (\hat{\bs{w}}_g - \bs{w}_g^*) \qquad\text{for}\quad 
\bar{\bs{p}} = \sum_{t=1}^T \bs{p}_t/T .
\]
As an indication of the sensitivity (or power) of the test, one can apply the test to 
\[
\Delta' = \widehat{\bar{P}}_g - \bar{P}_{g'}^* = \bar{\bs{p}}^{\top} (\hat{\bs{w}}_g - \bs{w}_{g'}^*) 
\]
for $g' \neq g$. Since the null hypothesis $E(\Delta') =0$ is known to be untrue \emph{a priori}, the $Z$-test \eqref{Ztest} is shown to lack sensitivity if it fails to reject such a null hypothesis.

\begin{table}[ht]
\centering
\caption{Results of $Z$-test with $(g,g')$ for mean source effect over 36 months.}
\begin{tabular}{|c|c|c|c|} \hline
Set-up & Observed Mean Effect & Test statistic & P-value \\ \hline  
$\Delta$ with $(g,g') = (4,4)$ &  0.001099 & 0.03803 & 0.970 \\  
$\Delta'$ with $(g,g') = (4,1)$ & 0.001059 & 0.03666 & 0.971 \\ \hline
\end{tabular} \label{tab-Ztest} \end{table}

Table \ref{tab-Ztest} gives the $Z$-test results using $\bar{\bs{p}}$ calculated over 36 months. In the case of $(g,g') = (4,4)$, we are testing the true effect of using the proxy weights that ideally should be close to the true weights. The p-value is close to 1, so the observed mean source effect over the whole period is very small compared to its uncertainty due to the sampling variance of the CES. Meanwhile, although $H_0:\, E(\Delta) =0$ cannot be rejected, neither is it well corroborated, because the test lacks sensitivity. As can be seen in Table \ref{tab-Ztest}, one cannot reject $E(\Delta') =0$ with $(g,g') = (4,1)$, where the $p$-value is about the same, despite one can detect in Figure \ref{fig-scat} that the source effect is different in the two cases.

Whilst the limited CES sample size is ultimately a cause for the insensitivity of the $Z$-test, one should recognise that the construction of the test statistic can be important. According to \eqref{source}, the observed source effect $\Delta$ is likely to be small, unless the relative weight difference $b_i$ tends to correlate with $p_i$. However, the empirical evidences are such that a covariation is not the case with these data. For instance, one can repeatedly apply the $Z$-test to $\Delta_t$ with $g=g'=4$, for $t = 1, ..., 36$, and the lowest p-value is 0.9038 over the 36 months. This suggests that $\Delta_t$ is always relatively small, given these $\bs{p}_t$'s. Moreover, on repeatedly applying the $Z$-test to $\Delta_t'$ with $(g,g') = (4,1)$ one obtains the lowest p-value 0.8098 over the 36 months. This shows that neither does $b_i' = \hat{w}_g/w_{g'}^* -1$ correlate appreciably with $p_i$, when the weights refer to quite different sub-populations.

\subsection{A unity slope test}

To construct a more sensitive test, we would like to make better use of the regression patterns seen in Figure \ref{fig-scat}. The idea is that the differences between any $\widehat{P}_t = \bs{p}_t^{\top} \hat{\bs{w}}$ and $P_t^* = \bs{p}_t^{\top} \bs{w}^*$ can become more pronounced over $t=1, ..., T$, if the discrepancies between the weights act to accentuate the different group-specific \emph{trends} of $\bs{p}_t$. To this end we now rewrite the collection of price indices $\{ p_{it}: i = 1, ..., m;  t = 1, ..., T\}$ as 
\begin{equation} \label{rewrite}
p_{it} = \bar{p}_i + \gamma_i \delta_t + e_{it}
\end{equation}
where $\bar{p}_i = \sum_{t=1}^T p_{it}/T$ is the mean price index for the $i$th group, and $\delta_t$ is the distance (in time) from $t$ to the mean time point such that $\sum_{t=1}^{T} \delta_t \equiv 0$, and $\gamma_i$ is the group-specific trend, and $e_{it}$ is a residual term satisfying $\sum_{t=1}^T e_{it} \equiv 0$. We emphasise that \eqref{rewrite} is not a statistical model of $p_{it}$, but simply re-expresses $\{ p_{it}\}$ in terms of $\bar{p}_i$, $\gamma_i$, $\delta_t$'s and $e_{it}$'s. 

Now, given any age-group weights $\hat{\bs{w}}_g$ and $\bs{w}_{g'}^*$, we have 
\begin{align*}
& P_{g't}^* = \bs{p}_t^{\top} \bs{w}_{g'}^* = \bar{P}_{g'}^* + \gamma_{g'}^* \delta_t + e_{g't}^* \quad\Leftrightarrow\quad 
\delta_t = (P_{g't}^* - \bar{P}_{g'}^* - e_{g't}^*) / \gamma_{g'}^* \\
& \widehat{P}_{gt} = \bs{p}_t^{\top} \hat{\bs{w}}_g = \widehat{\bar{P}}_g + \hat{\gamma}_g \delta_t + \hat{e}_{gt} 
\end{align*}
where $\bar{P}_{g'}^* =\sum_{i=1}^m w_{g'i}^* \bar{p}_i$, $\gamma_{g'}^* = \sum_{i=1}^m w_{g'i}^* \gamma_i$, $e_{g't}^* = \sum_{i=1}^m w_{g'i}^* e_{it}$, and $\widehat{\bar{P}}_g =\sum_{i=1}^m \hat{w}_{gi} \bar{p}_i$, $\hat{\gamma}_g = \sum_{i=1}^m \hat{w}_{gi} \gamma_i$, $\hat{e}_{gt} = \sum_{i=1}^m \hat{w}_{gi} e_{it}$. Substituting $\delta_t$ from the first equation above into the second one, we obtain
\[
\widehat{P}_{gt} = \big( \widehat{\bar{P}}_g - \frac{\hat{\gamma}_g}{\gamma_{g'}^*} \bar{P}_{g'}^* \big) 
+ \frac{\hat{\gamma}_g}{\gamma_{g'}^*} P_{g't}^* + \big( \hat{e}_{gt} - \frac{\hat{\gamma}_g}{\gamma_{g'}^*} e_{g't}^* \big) ~.
\]
This explains all the linear relationships seen in Figure \ref{fig-scat}, between $\hat{P}_{gt}$ and $P_{g't}^*$ for different combinations of $(g,g')$. In particular, we are interested in testing whether $\gamma_g = \gamma_g^*$, e.g. in the case of $g=g'=4$, where $\gamma_g = E(\hat{\gamma}_g)$ is based on the true weights $\bs{w}= E(\hat{\bs{w}})$, i.e. 
\[
H_0:~ E(\hat{\gamma}_g) = \gamma_g^* ~ \text{ vs. }  H_1:~ E(\hat{\gamma}_g) \neq \gamma_g^* .
\]

Now that the true slope of regressing $\widehat{P}_{gt}$ on $P_{gt}^*$ is 1 under the null hypothesis above, let $\hat{\beta}$ be the ordinary least square (OLS) fit, given by
\[
\hat{\beta} = \frac{\sum_{t=1}^T (P_{gt}^* - \bar{P}_g^*) (\widehat{P}_{gt} - \widehat{\bar{P}}_g)}{\sum_{t=1}^T (P_{gt}^* - \bar{P}_g^*)^2} 
= \bs{d}^{\top} \hat{\bs{w}}_g 
\]
where 
\[
\bs{d} = \sum_{t=1}^T (P_{gt}^* - \bar{P}_g^*)(\bs{p}_t - \bar{\bs{p}}) / \sum_{t=1}^T (P_{gt}^* - \bar{P}_g^*)^2 ~.
\]
The null hypothesis $E(\hat{\gamma}_g) = \gamma_g$ is then the same as that of the \emph{unity slope}, i.e. 
\begin{equation} \label{H0H1slope}
H_0:~ E(\hat{\beta}) = 1 ~ \text{ vs. }  H_1:~ E(\hat{\beta}) \neq 1 ~.
\end{equation}
A test statistic, to be referred to as the $B$-test, for \eqref{H0H1slope} is 
\begin{equation} \label{Btest}
B = (\hat{\beta} -1)/ \sqrt{V(\hat{\beta})} ~\sim~ N(0,1) \qquad\text{where}\quad   
V(\hat{\beta}) = \bs{d}^{\top} V(\hat{\bs{w}}_g) \bs{d} ~.
\end{equation}
As an indication of the sensitivity of the test, one can apply \eqref{Btest} to the OLS fit of $\widehat{P}_{gt}$ on $P_{g't}^*$, where $g \neq g'$ and the null hypothesis of unity slope is unlikely to hold, i.e. using
\[
\hat{\beta}' = \bs{d}'^{\top} \hat{\bs{w}}_g \qquad\text{for}\quad 
\bs{d}' = \sum_{t=1}^T (P_{g't}^* - \bar{P}_{g'}^*)(\bs{p}_t - \bar{\bs{p}}) / \sum_{t=1}^T (P_{g't}^* - \bar{P}_{g'}^*)^2 ~.
\] 

\begin{table}[ht]
\centering
\caption{Results of $B$-test with $(g,g')$ for unity slope over 36 months.}
\begin{tabular}{|c|c|c|c|} \hline
Set-up & OLS fit & Test statistic & P-value \\ \hline  
$\hat{\beta}$ with $(g,g') = (4,4)$ &  1.0065 & 0.2120 & 0.832 \\  
$\hat{\beta}'$ with $(g,g') = (4,1)$ & 1.0861 & 2.6122 & 0.009 \\
$\hat{\beta}$ with $(g,g') = (1,1)$ & 0.9886 & -0.1783 & 0.859 \\ \hline
\end{tabular} \label{tab-Btest} \end{table}

Table \ref{tab-Btest} gives three $B$-test results. Again, in the case of $(g,g') = (4,4)$, we are testing the true effect of using the proxy weights that ideally should be close to the true weights, now using the $B$-test instead of the $Z$-test.  The p-value is 0.832, and the OLS fit deviates little to the unity slope, compared to its uncertainty due to the sampling variance of the CES. Similarly in the case of $(g,g') = (1,1)$. Unlike with the $Z$-test, the results here provide stronger corroboration to these proxy CPI weights, because the p-value of the $B$-test is 0.009 based on $\hat{\beta}'$ with $(g,g') =(4,1)$, which is significant at the $5\%$ level. In other words, it is sensitive and one is able to reject this untrue null hypothesis.

\section{Accuracy of proxy-weight index} \label{uncertainty}

In reality the true source effect will not be zero, as long as the proxy weights $\bs{w}_g^*$ are not exactly equal to the true weights $\bs{w}_g$. To address question-II raised in the Introduction, we will treat $\widehat{P}_{gt}$ as an unbiased audit sample estimator of $P_{gt}$, so that the difference $P_{gt}^* - \widehat{P}_{gt}$ is an unbiased estimator of the bias of $P_{gt}^*$ based on the proxy weights. Assume zero variance of $P_{gt}^*$. Let $\widehat{V}(\widehat{P}_{gt})$ be an unbiased estimator of the variance of $\widehat{P}_{gt}$. An unbiased estimator of the MSE of $P_{gt}^*$ can be given by 
\begin{equation} \label{mse}
\mbox{mse}(P_{gt}^*) = (P_{gt}^* - \widehat{P}_{gt})^2 - \widehat{V}(\widehat{P}_{gt}) ~.
\end{equation}
However, for the data of this study, the MSE estimate is negative in \emph{all} the 36 months, due to the relatively large sampling variance of $\widehat{P}_{gt}$ , as evidenced from the Z-test results reported earlier. This reveals an important drawback of using MSE as the uncertainty measure in the present setting: unless the audit sample is sufficiently large, indeed much larger than the actual CES, unbiased MSE estimation may fail to yield any meaningful uncertainty measure, when the proxy-weight index has a relatively small bias.   

Below we shall propose a new accuracy measure, discuss its properties, and demonstrate empirically the advantages of this novel proposal for big data price index.

\subsection{Evaluation coverage (I)}

The concept of \emph{evaluation coverage} as an accuracy measure is as follows. Let $\theta_0$ be a target scalar parameter value. Let $A_{\alpha}$ be an imaginary autonomous \emph{evaluation} confidence interval,  which is centred at $\theta_0$ and covers it with a probability $\alpha$ of choice, say $\alpha = 0.95$. Let $\theta^*$ be a constant value in the parameter space. To assess how good $\theta^*$ is as an estimate of $\theta_0$, we now calculate the probability that $\theta^*$ is covered by $A_{\alpha}$. We shall call this probability $c(\theta) = \mbox{Pr}(\theta^* \in A_{\alpha})$ the \emph{evaluation coverage} of $\theta^*$ by $A_{\alpha}$. 

Suppose $c(\theta_0) = 0.95$ and $c(\theta^*) = 0.93$, which means the confidence interval $A_{\alpha}$ that covers $\theta_0$ in 95\% of the times would cover the estimate $\theta^*$ in 93\% of the times. In other words, if one treats $\theta^*$ as the true value, then the $95\%$ confidence interval $A_{\alpha}$ would fail to cover it only $2\%$ more often than when one correctly holds $\theta_0$ to be the truth. As will be shown later, the evaluation coverage of a constant $\theta^*$ by $A_{\alpha}$ reaches its maximum value $\alpha$ if $\theta^* = \theta_0$, otherwise it decreases as $|\theta^* - \theta_0|$ increases. In this way, the evaluation coverage ratio $c(\theta^*)/c(\theta_0)$ can provide a measure of the accuracy of $\theta^*$, which achieves its maximum value 1 when $\theta^* = \theta_0$, and decreases monotonously to 0 as the bias of $\theta^*$ increases to infinity. This is appropriate when $\theta^*$ is based on big data, such that it is associated with a non-negligible bias but a negligible variance for all practical purposes.

Moreover, let $\hat{\theta}_n$ be an unbiased estimator of $\theta_0$ based on a probability sample of size $n$, such that its variance decreases as $n$ increases. One can compare the big data estimate $\theta^*$ and the finite-sample estimator $\hat{\theta}_n$ according to their respective evaluation coverages, where $c(\hat{\theta}_n) = \mbox{Pr}(\hat{\theta}_n \in A_{\alpha})$. One may consider $\theta^*$ to be better than $\hat{\theta}_n$ if $c(\theta^*) > c(\hat{\theta}_n)$, and vice versa. Indeed, the sample size $n$ at which $c(\theta^*) = c(\hat{\theta}_n)$ can be considered the break-even point, which indicates the cost that is required if one opts to estimate $\theta_0$ based on a designed probability sample, instead of based on the available big data.

In principle one can of course make the comparison between $\theta^*$ and $\hat{\theta}_n$ based on their respective MSEs instead. In practice, though, one needs to estimate the MSE, which can be particularly difficult for $\theta^*$, if it requires a very large (hence costly) probability audit sample, as noticed previously in connection with \eqref{mse}. In contrast, as we will explain and demonstrate below, one can employ a much smaller audit sample, in order to estimate the evaluation coverage of $\theta^*$ as its accuracy measure, denoted by $\hat{c}(\theta^*)$, and obtain readily a meaningful comparison between $\hat{c}(\theta^*)$ and $\hat{c}(\hat{\theta}_n)$. This is an important advantage for adopting the evaluation coverage as an accuracy measure instead of the MSE.

\subsection{Evaluation coverage (II)} 

Here we define the evaluation coverage and describe its properties in more precise terms. Let $A_{\alpha}$ be a $100\alpha\%$ autonomous normal \emph{evaluation} confidence interval of $\theta_0$, given by
\[
A_{\alpha} = (Z_A - \omega,~ Z_A + \omega) ~,
\]
which is of the width $2\omega$, where
\[
Z_A - \theta_0 \sim N(0, \sigma_{\alpha, \omega}^2) \qquad\text{and}\qquad \sigma_{\alpha,\omega} = \omega/\kappa_{\alpha}
\]
and $\kappa_{\alpha}$ is the $(1+\alpha)/2$ quantile of $N(0,1)$. The interval $A_{\alpha}$ is said to be autonomous, because it is an imaginary confidence interval, independent of any actual observations available or collected to estimate $\theta_0$. Let $\hat{\theta}$ be an estimator of $\theta_0$, for which one is interested to obtain an accuracy measure using $A_{\alpha}$, and $\hat{\theta}$ is statistically independent of $A_{\alpha}$ by definition. The evaluation coverage of $\hat{\theta}$ is the probability that it is covered by $A_{\alpha}$, which is given by
\begin{equation}\label{acvr}
c(\hat{\theta}) = \mbox{Pr}(\hat{\theta} \in A_{\alpha}) ~.
\end{equation}

Notice that we have $c(\theta_0) = \alpha$, i.e., the evaluation coverage of the true parameter value $\theta_0$ is $\alpha$ by definition. Moreover, the evaluation coverage of the same estimator $\hat{\theta}$ generally varies with the choice of $\omega$, which determines the width of $A_{\alpha}$. The choice of $\omega$ affects the stringency of evaluation. For instance, if one chooses $\omega =0.01$ as a price index precision requirement \emph{a priori}, then the interval $A_{\alpha}$ will correspond to an evaluation precision of $\pm 1\%$ on either side of the true index. One can relate the choice to a normally distributed random variable that is centred on the true value $\theta_0$, with a standard deviation $\sigma_{\alpha,\omega} = \omega/\kappa_{\alpha}$. For instance, in the context of CPI, one can set $\omega$ in correspondence to the sampling variance of the CES-based weights $\hat{\bs{w}}$, in which case the accuracy of the proxy weights $\bs{w}^*$ will be measured against a `yardstick' that can be related to the tangible precision of the CES. The approach will be illustrated in Section \ref{appl}.

Some properties of the evaluation coverage as an uncertainty measure are given below as Results 1 - 3, the proofs of which are given in Appendix \ref{proofs}.

\paragraph{Result 1} For any constant $\theta^*$ in the parameter space, we have $0< c(\hat{\theta}) \leq \alpha$, where the maximum is attained iff $\theta^* = \theta_0$. 

\bigskip
It follows that the further away a zero-variance point estimator is from $\theta_0$, the lower is its evaluation coverage, regardless the choice of $(\alpha, \omega)$. For any two $\theta^* \neq \theta'$ in the parameter space, if $|\theta' - \theta_0 | > | \theta^* - \theta_0 |$, then $c(\theta') < c(\theta^*)$.

\paragraph{Result 2} If  $\hat{\theta}^* \sim N(\theta^*, \tau^2)$ where $\theta^* \in (\theta_0 - \omega, \theta_0 + \omega)$, then $c(\hat{\theta}^*) < c(\theta^*)$.

\bigskip
In other words, extra variance reduces the evaluation coverage, as long as the absolute bias is less than $\omega$, i.e., $\theta^*$ is not too far away from $\theta_0$. Moreover, if $\theta^* \in (\theta_0 -\omega, \theta_0 + \omega)$, and $\hat{\theta}_1^* \sim N(\theta^*, \tau_1^2)$ and $\hat{\theta}_2^* \sim N(\theta^*, \tau_2^2)$ where $\tau_1 < \tau_2$, then $c(\hat{\theta}_1^*) < c(\hat{\theta}_2^*)$. However, notice that extra variance could increase the evaluation coverage if the bias is sufficiently large. To see why the latter is the case, let $\theta^*$ be so far away from $\theta_0$ that its evaluation coverage is virtually zero, then an normally distributed estimator, which is centred on $\theta^*$ but has a large variance, can actually be much closer to $\theta_0$ from time to time, which increases its chance of being covered by the evaluation confidence interval. In contrast, using MSE as the criterion, the latter estimator would always be worse due to the extra variance. This is an example of the difference between the two uncertainty measures. 

\paragraph{Result 3} If $\hat{\theta}^* \sim N(\theta^*, \tau^2)$ and $\hat{\theta}' \sim N(\theta', \tau^2)$, then $c(\hat{\theta}') < c(\hat{\theta}^*)$ if $|\theta' - \theta_0 | > | \theta^* - \theta_0 |$.

\bigskip
Result 3 is complementary to Result 2, that is, given two estimators with the same variance, the one with less bias has a higher evaluation coverage.

The evaluation coverage can thus communicate in an intuitive and meaningful way the accuracy of any estimator, including a zero-variance big data estimate. Of course, in practice, one needs  to estimate the evaluation coverage. Let $\theta^*$ be a constant in the parameter space, we have
\[
c(\theta^*) = \mbox{Pr}\big( \frac{\theta^* -\theta_0}{\sigma_{\alpha,\omega}} - \kappa_{\alpha} \leq 
\frac{Z_A -\theta_0}{\sigma_{\alpha,\omega}} \leq \frac{\theta^* -\theta_0}{\sigma_{\alpha,\omega}} + \kappa_{\alpha} \big) ~,
\] 
where the only unknown quantity is $\theta_0$. Given a point estimate, denoted by $\dot{\theta}_0$, we obtain 
\[
\hat{c}(\theta^*) =\mbox{Pr}\big( \frac{\theta^* -\theta_0}{\sigma_{\alpha,\omega}} - \kappa_{\alpha} \leq 
\frac{Z_A - \dot{\theta}_0}{\sigma_{\alpha,\omega}} \leq \frac{\theta^* -\theta_0}{\sigma_{\alpha,\omega}} + \kappa_{\alpha} \big) ~.
\] 
It is thus clear that one only needs the point estimate $\dot{\theta}_0$ derived from an audit sample, in order to estimate the evaluation coverage of $\theta^*$. The variance of $\dot{\theta}_0$ is not needed directly, which affects only the variance of $\hat{c}(\theta^*)$. This allows one to obtain a meaningful accuracy measure of $\theta^*$ using a much smaller audit sample, without being critically constrained by the variance of audit sampling. In contrast, the variance of $\dot{\theta}_0$ affects the MSE estimate of $\theta^*$ directly, i.e. $\mbox{mse}(\theta^*) = (\theta^* - \dot{\theta}_0)^2 - \widehat{V}(\dot{\theta}_0)$, in addition to the variance of $\mbox{mse}(\theta^*)$.  

Meanwhile, let $\hat{\theta}$ be an unbiased estimator of $\theta_0$. Due to the independence between $\hat{\theta}$ and $Z_A$, where $A_{\alpha} = (Z_A- \omega, Z_A +\omega)$ and $E(Z_A) = \theta_0$ by definition, we have 
\[
\hat{\theta} - Z_A \sim N(0, \tau^2) \qquad\text{where}\quad \tau^2 = \sigma_{\alpha, \omega}^2 + V(\hat{\theta}) ~,
\]
as long as $\hat{\theta}$ can be assumed to have the normal distribution $N\big(\theta_0, V(\hat{\theta})\big)$.
It follows that the evaluation coverage of $\hat{\theta}$ can be given as
\[
c(\hat{\theta}) = \mbox{Pr}\big( -\omega \leq \hat{\theta} - Z_a \leq \omega) ~,
\]
which can be calculated without actually collecting the data that is needed to produce $\hat{\theta}$. All one needs is to stipulate the variance $V(\hat{\theta})$, in order to conclude how $\theta^*$ would have compared to an unbiased estimator $\hat{\theta}$ that is of the precision specified in terms of $V(\hat{\theta})$.

\subsection{Application to food index} \label{appl}

Let us now apply the evaluation coverage to the food price index based on proxy weights, which allows us to appreciate its property empirically. Let $P_t^* = \bs{p}_t^{\top} \bs{w}^*$ be the index for month $t$. The evaluation coverage of $P_t^*$ by $A_{\alpha}$ is given by \eqref{acvr}, 
\[
c^* := c(P_t^*) = \mbox{Pr}(P_t^* - \omega \leq Z_A \leq P_t^* + \omega) 
= \Phi\big( \frac{P_t^* - P_t}{\sigma_{\alpha,\omega}} + \kappa_{\alpha} \big) 
-  \Phi\big( \frac{P_t^* - P_t}{\sigma_{\alpha,\omega}} - \kappa_{\alpha} \big) 
\]
where $P_t = \bs{p}_t^{\top} \bs{w}$ is the true index for month $t$, and $\Phi$ denotes the cumulative distribution function of standard normal distribution. Based on an audit sample, we obtain $\widehat{P}_t^A = \bs{p}_t^{\top} \hat{\bs{w}}_A$, where $\hat{\bs{w}}_A$ are the CPI weights estimated from the audit sample, and the estimator
\[
\hat{c}^* = \Phi\big( \frac{P_t^* - \widehat{P}_t^A}{\sigma_{\alpha,\omega}} + \kappa_{\alpha} \big) 
-  \Phi\big( \frac{P_t^* - \widehat{P}_t^A}{\sigma_{\alpha,\omega}} - \kappa_{\alpha} \big) 
\]
The variance of $\hat{c}^*$ can be approximated as follows:
\begin{align}
& \hat{c}^* \approx c^* + \sigma_{\alpha,\omega}^{-1} 
\Big( \phi\big( \frac{P_t^* - \widehat{P}_t^A}{\sigma_{\alpha,\omega}} + \kappa_{\alpha} \big) 
-  \phi\big( \frac{P_t^* - \widehat{P}_t^A}{\sigma_{\alpha,\omega}} - \kappa_{\alpha} \big) \Big) (\widehat{P}_t^A - P_t) \notag \\
& V( \hat{c}^*) \approx \frac{V(\widehat{P}_t^A)}{\sigma_{\alpha,\omega}^2}
\Big( \phi\big( \frac{P_t^* - \widehat{P}_t^A}{\sigma_{\alpha,\omega}} + \kappa_{\alpha} \big) 
-  \phi\big( \frac{P_t^* - \widehat{P}_t^A}{\sigma_{\alpha,\omega}} - \kappa_{\alpha} \big) \Big)^2 \label{vstar}
\end{align}
provided unbiased audit-sample estimator $\widehat{P}_t^A$, where $V(\widehat{P}_t^A) = \bs{p}_t V(\hat{\bs{w}}_A) \bs{p}_t^{\top}$, and $\phi$ denotes the probability density function of standard normal distribution. Notice that only the audit-sample estimate $\widehat{P}_t^A$ is needed to estimate the evaluation coverage $c(P_t^*)$, whereas its variance $V(\widehat{P}_t^A)$ is needed to estimate the variance of $\hat{c}(P_t^*)$.

% figure floated here
\begin{figure}[H] 
\resizebox{163mm}{96mm}{\includegraphics{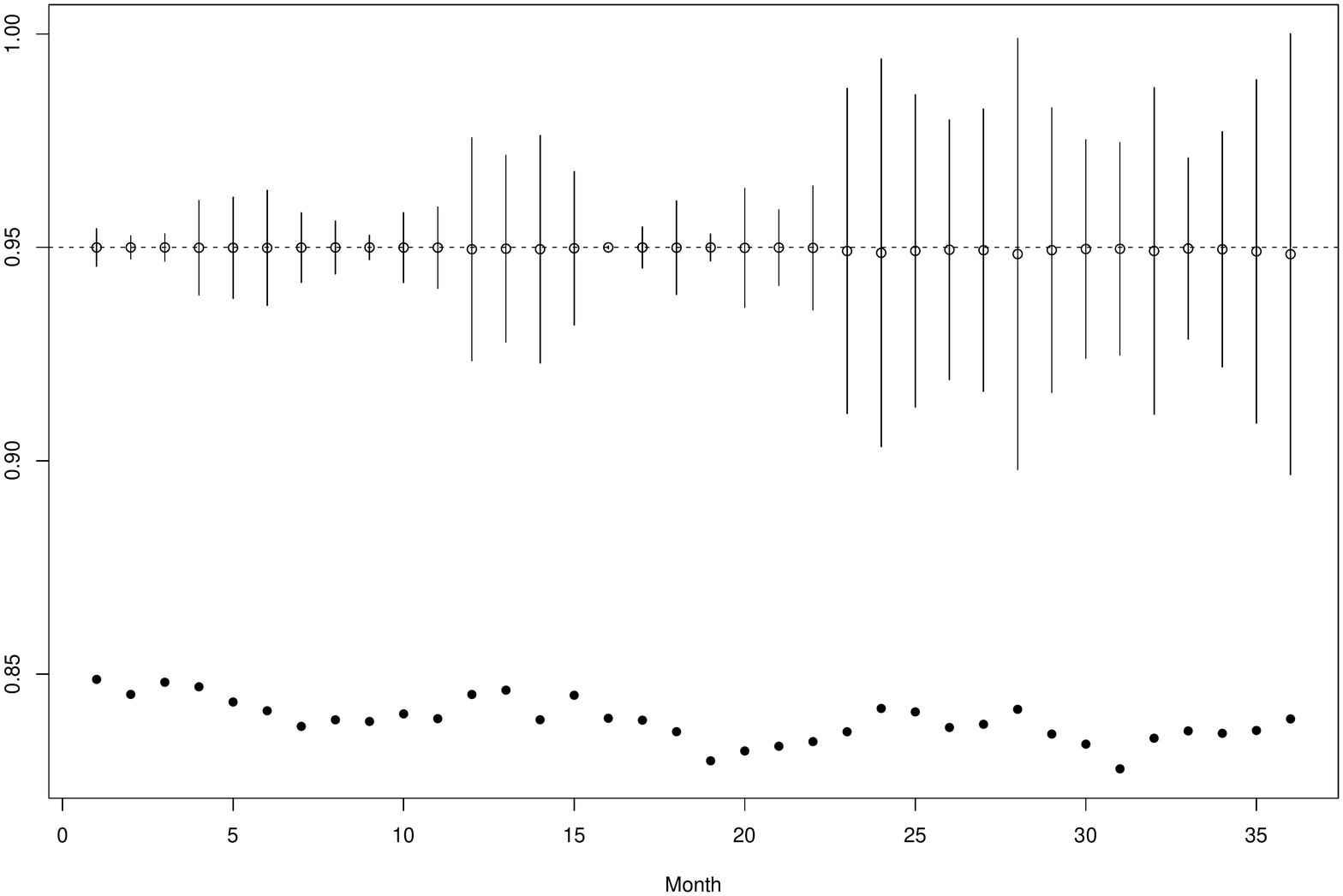}}
\resizebox{163mm}{96mm}{\includegraphics{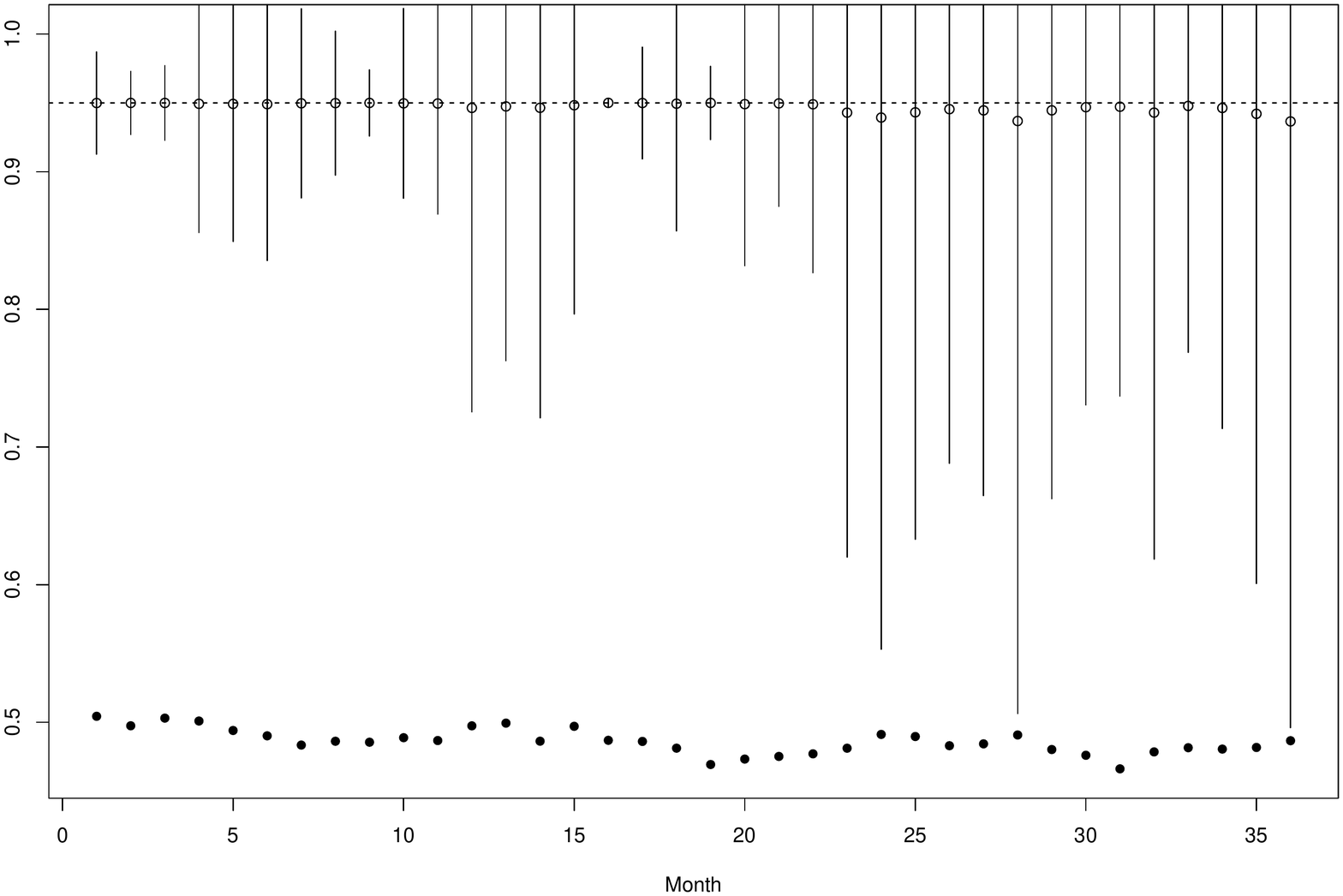}}
\caption{Evaluation coverage by $A_{0.95}$ of food index for age-group 4 over 36 months: $\hat{c}(\widehat{P}_{gt}^*)$ of proxy-weights index (circle), 95\% confidence interval of $\hat{c}(\widehat{P}_{gt}^*)$ marked by vertical line, $\hat{c}(\widehat{P}_{gt})$ of CES-weights index (solid), $c(P_{gt}) = 0.95$ of true index (horizontal dashed line). Top: $\omega = 2 \cdot se$ and $se = $ estimated average standard error of $\widehat{P}_{gt}$; bottom: $\omega = 0.02$.} \label{fig-coverage}
\end{figure}

Let $\widehat{P}_t = \bs{p}_t^{\top} \hat{\bs{w}}$ be the index based on the CES weights for month $t$. The evaluation coverage of $\widehat{P}_t$ by the same $A_{\alpha}$ above is given by
\[
c_s := c(\widehat{P}_t) = \mbox{Pr}(-\omega \leq Z_A - \widehat{P}_t \leq \omega) = \Phi(\omega/\tau) - \Phi(-\omega/\tau) 
\]
assuming the CES-based $\widehat{P}_t$ is unbiased for $P_t$, where $\tau^2 = \sigma_{\alpha,\omega}^2 + V(\widehat{P}_t)$. Notice that,  Given an estimator $\hat{v} = \widehat{V}(\widehat{P}_t^A)$, and $\hat{\tau}^2 = \sigma_{\alpha,\omega}^2 + \hat{v}$, one obtains $\hat{c}_s$ and its approximate variance as
\begin{align*}
& \hat{c}_s =  \Phi(\omega/\hat{\tau}) - \Phi(-\omega/\hat{\tau}) \\
& \widehat{V}(\hat{c}_s) \approx \big( \phi(\omega/\hat{\tau}) + \phi(-\omega/\hat{\tau}) \big)^2 
\big( \omega^2/4\hat{\tau}^2 \big) \widehat{V}(\hat{v})
\end{align*}
where $-\omega/2\hat{\tau}^3$ is an estimate of $\partial (\omega/\hat{\tau})/\partial \hat{v}$ and $\widehat{V}(\hat{v})$ an estimate of $V(\hat{v})$.

\begin{table}[ht]
\begin{center}
\caption{Evaluation coverage by $A_{0.95}$ of index by proxy or CES weights over 36 months.}
\begin{tabular}{l | cc | cc} \hline
& \multicolumn{2}{c|}{$\omega = 0.058$} & \multicolumn{2}{c}{$\omega = 0.02$} \\  
Summary & Proxy weights & CES-equivalent & Proxy weights & CES-equivalent \\ \hline
Minimum  & 0.948  & 0.828 & 0.937 & 0.466 \\ 
First Quantile & 0.949  & 0.836 & 0.945 & 0.481 \\
 Median & 0.950 & 0.839 & 0.949 & 0.486 \\
 Mean  & 0.950  & 0.839 & 0.947 & 0.486\\
 Third Quantile & 0.950 & 0.842 & 0.950 & 0.491 \\
 Maximum & 0.950 & 0.849 & 0.950 & 0.504 \\ \hline
\end{tabular} \label{tab-coverage} \end{center} \end{table}

Fixing $\alpha$ at 0.95, we can vary the stringency of evaluation in terms of $\omega$. The average estimated standard error of the CES-based index $\widehat{P}_{4t}$ is $se = 0.029$ over the 36 months, for the 4th age group $g=4$. As the first choice we set $\omega = 2\cdot se$, in which case $A_{0.95}$ can be considered as a 95\% confidence interval of the true index $P_{4t}$, based on an audit sample of the same precision as the CES. The estimated evaluation coverage $\hat{c}(P_{4t}^*)$ is evaluated at $\dot{P}_{4t} = \widehat{P}_{4t}$, using the CES as an audit sample, which tells us how often the proxy-weights index $P_{4t}^*$ is covered by the same $A_{0.95}$. In addition, we calculate the evaluation coverage of an unbiased index that is of the same variance as the CES-based index.  The results are shown in the top panel of Figure \ref{fig-coverage} and summarised in Table \ref{tab-coverage}. 

The evaluation coverage of $P_{4t}^*$ varies from 0.948 to 0.950 over 36 months, which is very close to $c(P_{4t}) = 0.95$ of the true index. They compare favourably to the evaluation coverage of an CES-equivalent unbiased index, which varies from 0.829 to 0.849. The confidence interval of $c(P_{4t}^*)$ are marked by vertical lines in Figure \ref{fig-coverage}, which is seen to increase gradually in width over 2015 - 2017, before reaching a different level for the 12 months in 2017. Still, the evaluation coverage of an CES-equivalent unbiased index is out of these confidence intervals throughout the whole period. Altogether the results suggest that the bias of $P_{4t}^*$ is small enough for it to outperform the CES-based index. Indeed, it turns out that the standard error of an unbiased index needs to be reduced to about 1/9 of that of the CES-based index, in order to achieve about the same median and mean evaluation coverages as the proxy index $P_{4t}^*$ over the 36 months. But such an increase of the CES sample size is unthinkable in reality due to the associated cost.

Next, the results are given in Figure \ref{fig-coverage} and Table \ref{tab-coverage}, where $\omega$ is reduced to 0.02. The width of $A_{0.95}$ is reduced to about 1/3 of that used above. The same bias of $P_{4t}^*$ leads then to a lower evaluation coverage for the same level $\alpha = 0.95$, which varies now from 0.937 to 0.950 over the three-year period. However, the reduction of evaluation coverage is much greater for an unbiased index of the same variance as the CES-based index, which varies from 0.466 to 0.504. In other words, subjected to the increased stringency of evaluation, the proxy index $P_{4t}^*$ compares even more favourably to the CES-based index. 

Finally, Figure \ref{fig-coverage} shows that the width of the confidence interval of $c(P_{4t}^*)$ increases quite fast as $\omega$ is reduced from 0.058 to 0.02. The reason is clear from \eqref{vstar}. While the variance due to audit sampling $V(\widehat{P}_{4t})$ remains the same, the other two terms are affected by the change in $\omega$:  in the denominator $\sigma_{\alpha, \omega} = \omega/\kappa_{\alpha}$ is reduced proportionally with $\omega$, in the numerator the term $\phi\big( \frac{P_{4t}^* - \widehat{P}_{4t}^A}{\sigma_{\alpha,\omega}} + \kappa_{\alpha} \big) -  \phi\big( \frac{P_{4t}^* - \widehat{P}_{4t}^A}{\sigma_{\alpha,\omega}} - \kappa_{\alpha} \big) $ is increased, because the reduction of $\sigma_{\alpha, \omega}$ increases the asymmetry of $\frac{P_{4t}^* - \widehat{P}_{4t}^A}{\sigma_{\alpha,\omega}} \pm \kappa_{\alpha}$ around 0. The two effects amplify each other, increasing $\sqrt{V(\hat{c}^*)}$ more quickly than a rate proportional to $0.058/0.02$.

\section{Final remarks} \label{final}

In the above we have developed an audit sampling approach for big data statistics, which can be classified as privacy-preserving as it does not require linking the data at the individual level. It is shown that testing the bias of a zero-variance big data estimate may require careful reasoning, in order to achieve sufficient power of test. A difficult challenge arises in situations, where the audit sampling variance is relatively large compared to the bias of big data estimate, which results in a negative (hence unusable) MSE estimate. We develop the evaluation coverage as a novel accuracy measure, which requires only the point estimate derived form the audit sample and is therefore not constrained by small audit sample sizes. This provides one readily with the means to assess the big data bias against the cost and burden of idealistically speaking unbiased estimation based on traditional survey sampling. The evaluation coverage is flexible to apply, where one can either use an existing survey on the same topic or, if such a survey does not exist, undertake separate agile audit sampling with relatively small sample size and low cost.

\appendix
 
\section{Proof of Result 1, 2 and 3} \label{proofs}

Let $\theta_0$ be the true scalar parameter value. Let $Z\sim N(\theta_0, \sigma^2)$ be a normally distributed random variable. The 
shortest $100\alpha\%$ confidence interval of $\theta_0$ is $A_{\alpha,\omega} = Z \pm \omega$ with $\omega = \kappa_{\alpha} \sigma$, where $\sigma$ is a short-hand for $\sigma_{\alpha,\omega}$, and $c(\theta_0) = \mbox{Pr}(\theta_0 \in A_{\alpha,\omega}) = \alpha$, and $\kappa_{\alpha}$ is the $(1+\alpha)/2$ quantile of $N(0,1)$. Results 1 - 3 are given Proofs 1-3 below, respectively. 
%Let $\hat{\theta}^*$ be an estimator and let $\alpha(\hat{\theta}^*) = \mbox{Pr}\big( \hat{\theta}^* \in A_{\alpha,\omega} \big)$ given the choice of $\alpha$ and $\omega$. 

\bigskip \noindent 
\emph{Proof-1} Let $\Phi$ be the CDF of $N(0,1)$. Write $c^* = \alpha(\theta^*)$. We have
\begin{align*}
c^* & = \mbox{Pr}\big( \theta^* - \kappa_{\alpha} \sigma \leq Z \leq \theta^* + \kappa_{\alpha} \sigma \big) 
= \mbox{Pr}\big( \frac{\theta^* -\theta_0}{\sigma} - \kappa_{\alpha} \leq \frac{Z -\theta_0}{\sigma} 
\leq \frac{\theta^* -\theta_0}{\sigma} + \kappa_{\alpha} \big) \\
c^* - \alpha
& = \Phi(\kappa_{\alpha} + \frac{\theta^* -\theta_0}{\sigma}) - \Phi(-\kappa_{\alpha} + \frac{\theta^* -\theta_0}{\sigma})
- \Phi(\kappa_{\alpha}) + \Phi(-\kappa_{\alpha}) \\
& = \begin{cases}
\big( \Phi(\kappa_{\alpha} + \frac{\theta^* -\theta_0}{\sigma}) - \Phi(\kappa_{\alpha}) \big) - 
\big( \Phi(-\kappa_{\alpha} + \frac{\theta^* -\theta_0}{\sigma}) -  \Phi(-\kappa_{\alpha}) \big) < 0 & \text{if } \theta^* > \theta_0\\
0 & \text{if } \theta^* = \theta_0 \\
- \big( \Phi(\kappa_{\alpha}) - \Phi(\kappa_{\alpha} + \frac{\theta^* -\theta_0}{\sigma}) \big) +
\big( \Phi(-\kappa_{\alpha}) - \Phi(-\kappa_{\alpha} + \frac{\theta^* -\theta_0}{\sigma}) \big) < 0 & \text{if } \theta^* < \theta_0
\end{cases} \square
\end{align*}

\bigskip \noindent 
\emph{Proof-2} Let $\nu^2 = \sigma^2 + \tau^2$. Without losing generality, suppose $\theta^* \in [\theta_0, \theta_0 + \kappa_{\alpha} \sigma)$. We have
\begin{align*}
c(\hat{\theta}^*) & = \mbox{Pr}\big( - \kappa_{\alpha} \sigma \leq Z - \hat{\theta}^* \leq \kappa_{\alpha} \sigma \big) \\
& = \mbox{Pr}\left( - \frac{\kappa_{\alpha} \sigma}{\nu} + \frac{\theta^* -\theta_0}{\nu} 
\leq \frac{Z - \hat{\theta}^*}{\nu} + \frac{\theta^* - \theta_0}{\nu} 
\leq \frac{\kappa_{\alpha} \sigma}{\nu} + \frac{\theta^* -\theta_0}{\nu} \right) \\
& = \Phi(\frac{\kappa_{\alpha} \sigma}{\nu} + \frac{\theta^* -\theta_0}{\nu} ) 
- \Phi(- \frac{\kappa_{\alpha} \sigma}{\nu} + \frac{\theta^* -\theta_0}{\nu}) \\
&~~ < \Phi(\kappa_{\alpha} + \frac{\theta^* -\theta_0}{\sigma}) - \Phi(-\kappa_{\alpha} + \frac{\theta^* -\theta_0}{\sigma})
= c(\theta^*) 
\end{align*}
since $\kappa_{\alpha} \sigma+ (\theta^* -\theta_0) \geq 0$ and $-\kappa_{\alpha} \sigma + (\theta^* -\theta_0) \leq 0$. $\square$

\bigskip \noindent 
\emph{Proof-3} Let $\nu^2 = \sigma^2 + \tau^2$. We have
\begin{align*}
c(\hat{\theta}') - c(\hat{\theta}^*) & = \big[ \Phi(\frac{\kappa_{\alpha} \sigma}{\nu} + \frac{\theta' -\theta_0}{\nu}) - 
\Phi(\frac{\kappa_{\alpha} \sigma}{\nu} + \frac{\theta^* -\theta_0}{\nu}) \big] \\
& \quad - \big[ \Phi(- \frac{\kappa_{\alpha} \sigma}{\nu} + \frac{\theta' -\theta_0}{\nu}) -
\Phi(- \frac{\kappa_{\alpha} \sigma}{\nu} + \frac{\theta^* -\theta_0}{\nu}) \big]
\end{align*}
Due to symmetry $c(\hat{\theta}^*) = c\big( \hat{\theta}^* + (2\theta_0 - \theta^*)\big)$, so we only need to consider the situation of $\theta' > \theta^* > \theta_0$. Then, the interval $(-\frac{\kappa_{\alpha} \sigma}{\nu} + \frac{\theta^* -\theta_0}{\nu}, -\frac{\kappa_{\alpha} \sigma}{\nu} + \frac{\theta' -\theta_0}{\nu})$ is closer to 0 than $(\frac{\kappa_{\alpha} \sigma}{\nu} + \frac{\theta^* -\theta_0}{\nu}, \frac{\kappa_{\alpha} \sigma}{\nu} + \frac{\theta' -\theta_0}{\nu})$, where the two have the same length. The result follows from
\[
\Phi(- \frac{\kappa_{\alpha} \sigma}{\nu} + \frac{\theta' -\theta_0}{\nu}) -
\Phi(- \frac{\kappa_{\alpha} \sigma}{\nu} + \frac{\theta^* -\theta_0}{\nu}) >
\Phi(\frac{\kappa_{\alpha} \sigma}{\nu} + \frac{\theta' -\theta_0}{\nu}) - 
\Phi(\frac{\kappa_{\alpha} \sigma}{\nu} + \frac{\theta^* -\theta_0}{\nu}) > 0 \square
\]
%Similarly, in the case of $\theta' < \theta^* < \theta_0$, we have
%\[
%\Phi(\frac{\kappa_{\alpha} \sigma}{\nu} + \frac{\theta^* -\theta_0}{\nu}) - 
%\Phi(\frac{\kappa_{\alpha} \sigma}{\nu} + \frac{\theta' -\theta_0}{\nu}) > 
%\Phi(- \frac{\kappa_{\alpha} \sigma}{\nu} + \frac{\theta^* -\theta_0}{\nu}) -
%\Phi(- \frac{\kappa_{\alpha} \sigma}{\nu} + \frac{\theta' -\theta_0}{\nu}) > 0
%\]

\end{document}